\documentclass{WileyMSP-template}

\usepackage{lineno}
\usepackage{amsthm}
\usepackage{xcolor}
\usepackage{amssymb}
\usepackage{amsmath}
\usepackage{ragged2e}

\newtheorem{theorem}{Theorem}[section]

\newtheorem{lemma}[theorem]{Lemma}

\theoremstyle{remark}
\newtheorem*{remark}{Remark}

\begin{document}
\justifying

\pagestyle{fancy}
\rhead{\includegraphics[width=2.5cm]{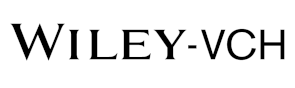}}

\title{On the large N convergence of the supermembrane matrix model}

\maketitle


\author{Mar\'ia del Pilar Garc\'ia del Moral}
\author{Pablo Le\'on*}
\author{Alvaro Restuccia}



\begin{affiliations}
M.P Garc\'ia del Pilar\\
Área de Física, Departamento de Química, Universidad de la Rioja, La Rioja 26006, Spain

P. Le\'on \\
Instituto de Física, Pontificia Universidad Católica de Chile, Av. Vicuña Mackenna 4860, Santiago, Chile.
Email Address: pleot@uc.cl

A. Restuccia \\
Departamento de Física, Universidad de Antofagasta, Universidad de Antofagasta, Campus Coloso, Aptdo 02800, Antofagasta, Chile

\end{affiliations}



\begin{abstract}

In this paper, the large N behavior of a supersymmetric matrix model is compared with its exact continuum description. We prove that the complete spectrum of a $SU(N)$ regularized  compactified M2-brane theory in the semiclassical approximation converges, in the large $N$ limit, to the exact spectrum in the same approximation.
To this end, we concentrate on the large $N$ limit of a supersymmetric matrix model describing a supermembrane with central charge on a toroidally compactified target space. We analyze, on the one hand, the supermembrane model formulated on a differentiable compact torus, with structure group given by the area-preserving diffeomorphisms, and, on the other hand, its associated regularized $SU(N)$ model. We emphasize in our analysis the structure of the local and global constraints of the regularized model, which generate the $SU(N)$ algebra and reproduce the area-preserving diffeomorphism algebra in the large N limit, together with the topological information associated with the central charge condition and the topology of the model. We explain the role of the central charge in the compactified supermembrane and how it allows a top-down $SU(N)$ regularization. It is known that its $SU(N)$ regularized model has a purely discrete spectrum with finite multiplicity. We prove that, in the semiclassical approximation, the spectrum of the Hamiltonian of the supersymmetric matrix model converges in the large-$N$ limit to that of the Hamiltonian of the supermembrane (M2-brane) with central charge.


\end{abstract}


\section{Introduction}
\label{introduction}

Matrix models provide non-perturbative methods that have been applied in a variety of contexts. Traditionally, they have been used to provide  rigorous analysis of quantum systems (see for example \cite{Halpern, Claudson, deWit2,dwln,deWit4, deWit3, Banks,IKKT, Dijkgraaf}).  In particular, they have been employed to characterize supersymmetric and non-supersymmetric matrix quantum-mechanical systems, as well as the slow-mode regime of Yang–Mills theories in arbitrary dimensions. In these frameworks, matrix models are capable of capturing non-perturbative features of QCD, including aspects related to the behavior of fluxtubes \cite{Mateos}.

A different, but equally important, modern application of matrix models pioneered in \cite{Banks} is the aim of understanding how geometry may emerge from an underlying quantum-mechanical system. This approach belongs to the broader context of emergent gravity and holography, as developed, for instance in \cite{Verlinde,Steinacker} (see also \cite{Minwalla, Myers, Berenstein}). From this perspective, matrix models provide a framework for an effective unification of Yang–Mills theory and gravity.

In this paper, we will follow the original line of work developed by  \cite{mpgm8} to  provide a consistent quantum description of M-Theory as a quantum gravity theory. It is based on the approach  \cite{deWit2,deWit4} and \cite{Banks}, but realized for a top-down regularization in target spaces with a compact sector.  In this sense, the matrix model may be understood as a matrix regularization of supermembrane theory, which is the source of eleven-dimensional supergravity and represents a fundamental M-brane in M-theory.

Finite-N matrix models are formulated in terms of finite-dimensional matrices acting on suitable Hilbert spaces. Following the original interpretation, presented in \cite{Hoppe,deWit2}, the large N limit of the spectral properties of the regularized theory is expected to reproduce the supersymmetric quantum spectrum of the M2-brane. Moreover, the existence of a non-trivial eigenvector associated with the ground state would describe the supergravity supermultiplet contained in the M2-brane spectrum. From the BFSS matrix model perspective \cite{Banks}, this construction was interpreted as describing M-theory in the infinite light-cone momentum frame, and therefore as providing a candidate formulation of quantum gravity. On the other hand, from the viewpoint of emergent geometry and holography, the large-N limit is expected to be related with the emergence of smooth 3+1-dimensional spacetime.

In all these approaches, the large-N limit of the finite-N matrix model is a central issue. However, this limit is not straightforwardly defined. Finite matrix models are starting points in some approaches usually obtained through an $SU(N)$ regularization, whereas in the large-N limit the corresponding symmetries should converge to the  $SU(\infty)$ \cite{Hoppe}, interpreted as the infinite-dimensional group of area-preserving diffeomorphisms (APD). Since different topologies are associated with inequivalent  area-preserving diffeomorphism groups, it is difficult to define rigorously the large-N limit of both the Hamiltonian, eigenvalues and the corresponding Fock spaces from a bottom-up approach, as well as to recover the complete topological information of the continuum theory.

Despite the relevance of this question, to the best of our knowledge there is no general result showing that the spectrum of a regularized theory can be consistently identified with the spectrum of the corresponding theory on the continuum, once the large-N limit has been taken. In this paper, we obtain explicit results for a concrete case. We restrict our analysis to the semiclassical formulation of the Supermembrane theory, since in this setting the spectrum of its Hamiltonian is known to be discrete and issues such as topology, geometry, and convergence can be handled consistently.
		
Our starting point is the supermembrane with central charge, which has been rigorously shown to possess a purely discrete quantum spectrum, in contrast with the supermembrane formulated in non-compact Minkowski spacetime. Moreover, the matrix regularization of Supermembrane theory on compactified target spaces from a top-down approach, is not generally defined. The difficulty arises from the regularization of the harmonic 1-forms associated with winding maps, which do not admit an $SU(N)$ regularization, as originally pointed out in \cite{deWit4,deWit3}. However, in \cite{mpgm8}, a consistent matrix model regularization for the supermembrane on a compactified  space was obtained when the theory is subject to a topological restriction known as the central charge condition, associated with an irreducible wrapping. This condition fixes the harmonic forms, which therefore cease to be dynamical degrees of freedom, allowing the theory to admit a consistent $SU(N)$ regularization. This regularization was used to study the spectrum at both the bosonic \cite{mpgm8} and supersymmetric \cite{mpgm11} levels, establishing that it is purely discrete, with eigenvalues of finite multiplicity.

This issue has often been bypassed by considering a bottom-up approach consisting in taking the matrix model, that is, the regularized description, as the fundamental theory, and then imposing isometries at the regularized level in order to mimic the Killing isometries of the continuum description \cite{Banks, Taylor, Myers, Steinacker2} or by including non commutative geometry \cite{Connes, Floratos}. However, it is important to emphasize that these two procedures  (top-down and bottom-up regularization) in general do not commute. In other words, regularizing the Supermembrane on a compact space is not equivalent to first considering the matrix model description of the Supermembrane and then imposing the isometries associated with the compact target space at the regularized level, unless harmonic sector is properly included. Consequently, the large-N limit of such a matrix model does not necessarily reproduce the original continuum theory. This provides an additional motivation for the analysis carried out in this work. Furthermore, we will explain that the central charge condition on the supermembrane toproidally compactified is necessary to obtain the $SU(N)$ matrix model, as originally showed in \cite{mpgm8.5}. To this end, we extend previous results on the large-$N$ limit of the semiclassical M2-brane to the supersymmetric case. We clarify the role of the global constraint present in the theory, an aspect that has not been addressed previously. More importantly, we show that convergence holds not only for individual eigenvalues but also for the full spectrum, including eigenvalue multiplicities.
	
The paper is organized as follows. In Section 2, we review the theory of the supermembrane with central charge. We describe its Hamiltonian formulation, the global area-preserving diffeomorphism constraints, and its expression in terms of the infinite-dimensional Fourier basis associated with the topology of the M2-brane. In Section 3, we present the corresponding matrix model regularization. In Section 4, we prove a new lemma that allows us to analyze the large-$N$ limit of the semiclassical spectrum of the supersymmetric matrix model and show that it coincides with the semiclassical spectrum of the supermembrane compactified on the same background and subject to the central charge condition. Finally, in Section 5, we present a brief discussion and our conclusions.


\section{The supermembrane with central charge}

The supermembrane theory extends the framework of string theory by allowing the fundamental object to carry one additional spatial dimension. In this formulation, the supermembrane is a $(2+1)$--dimensional dynamical surface propagating in an eleven--dimensional target spacetime. The covariant action was first introduced in~\cite{Bergshoeff} (see also~\cite{Bergshoeff3}) and is expressed in terms of the embedding coordinates
\begin{equation}
	Z^M(\xi) = \big(X^\mu(\xi),\, \theta^\alpha(\xi)\big),
\end{equation}
which depends on the worldvolume parameters $\xi^i$ $(i=0,1,2)$. It takes the form
\begin{equation}
	\label{supermembrane0}
	S[Z(\xi)] = - \int_{\mathbb{R}\times\Sigma} d^3\xi\,
	\left[
	\sqrt{-g(Z(\xi))} 
	+ \frac{1}{6}\,\epsilon^{ijk}\,
	\Pi_i^{A}\,\Pi_j^{B}\,\Pi_k^{C}\,C_{CBA}
	\right],
\end{equation}
where $\Sigma$ denotes the spatial two--dimensional Riemann surface swept out by the membrane. The induced metric on the worldvolume is defined by
\begin{equation}
	g_{ij} = \Pi_i^{a}\,\Pi_j^{b}\,\eta_{ab},
\end{equation}
with $\eta_{ab}$ the flat tangent--space metric and $\Pi_i^{A}$ the pullback of the superspace vielbein,
\begin{equation}
	\Pi_i^{A} = \partial_i Z^{M} E_{M}^{A}.
\end{equation}
Here $M=(\mu,\alpha)$ and $A=(a,\hat{\alpha})$ label the superspace and tangent superspace indices, respectively.

The supermembrane action becomes significantly simpler when considering a flat superspace, with or without compact directions, together with the light-cone gauge fixing. We denote the worldvolume coordinates as
\begin{equation}
	\xi^i=(t,\sigma^1,\sigma^2),
\end{equation}
where $(\sigma^1,\sigma^2)$ are local coordinates on the spatial Riemann surface $\Sigma$. The light-cone coordinates, so that $\mu=(+,-,\tilde{m})$ with $\tilde{m}=1,\dots,9$, are defined in terms of the eleven-dimensional target-space coordinates as
\begin{equation}
	X^{\pm}=\frac{1}{\sqrt{2}}\left(X^{10}\pm X^0\right),
\end{equation}
and analogously for the gamma matrices,
\begin{equation}
	\Gamma^{\pm}=\frac{1}{\sqrt{2}}\left(\Gamma^{10}\pm \Gamma^{0}\right).
\end{equation}
The remaining transverse maps are left unchanged. We then impose the light--cone gauge condition
\begin{equation}
	X^+ = t,
\end{equation}
together with the $\kappa$--symmetry fixing
\begin{equation}
	\Gamma^+\theta=0.
\end{equation}
The invariance under area--preserving diffeomorphisms is also used to partially
fix
\begin{equation}
	P_- = P_-^0\sqrt{W},
\end{equation}
where $\sqrt{W}$ is a scalar density satisfying
\begin{eqnarray}
	\int_{\Sigma}d^2\sigma\sqrt{W}=1.
\end{eqnarray}
Here $P_-^0$ denotes the zero mode of the light--cone momentum $P_-$ and is usually fixed to $P_-^0=1$ in the literature.
In order to describe a supermembrane sector with well-defined quantum properties, we restrict the theory to the toroidally compactified case subject to a non-trivial topological condition. This condition, usually refered in the literature as \textit{the central charge condition} \cite{Ovalle1}, is associated with an irreducible wrapping of the membrane over the compact target torus\footnote{By \textit{irreducible wrapping condition} it is meant a wrapping of the supermembrane over a compact (sub)manifold that cannot be unwrapped due to the existence of a topological invariant.}.

For the supermembrane with central charge we consider $M_9 \times T^2$, a simple but still non-trivial background where these effects can be easily observed, and take $\Sigma$ to be a genus one Riemann surface. Since we are considering compact directions, the embedding maps of the M2-brane from the worldvolume base $\Sigma$ into the target space split as follows:
\begin{eqnarray}
	X^r: \Sigma \rightarrow T^2, \quad X^m: \Sigma \rightarrow M_9,
\end{eqnarray}
with $r,s,t = 1,2$ and $m=3,\dots,9$. The maps $X^r$ obey the winding conditions
\begin{equation}
	\oint_{C_r}\!\big(dX^1 + i\,dX^2\big)
	= 2\pi R\,(l_s + m_s\,\tau)\,\delta^s{}_r, 
\end{equation}
where the torus $T^2$ is characterized by its complex modulus $\tau$ and radius $R$, and $C_r$ denotes a homology basis of the base manifold $\Sigma$. The
integer pairs $(l_s,m_s)$ denote the winding numbers, which may be arranged
into the matrix
\[
\mathbb{W} = \left(
\begin{array}{cc}
	l_1 & l_2 \\
	m_1 & m_2
\end{array} \right).
\]
For convenience, we introduce the complex 1-form
\[
dX \equiv dX^1 + i\,dX^2 .
\]

We consider the restriction to the topological sector with a non--vanishing central charge, defined by the following irreducible wrapping condition
\begin{equation}\label{CH}
	\int_{\Sigma} dX^r \wedge dX^s 
	= n\,\epsilon^{rs}\,(2\pi R)^2\operatorname{Im}(\tau),
	\qquad n \in \mathbb{Z}\setminus\{0\}.
\end{equation}
This condition signals the presence of a non--trivial $U(1)$ principal torus bundle over $\Sigma$, whose first Chern class is $c_1 = n$. It also imposes the irreducible wrapping condition
\begin{equation}
	\det\mathbb{W} = n,
\end{equation}
as a restriction on the winding matrix.

The 1-form associated with the membrane embedding along the torus directions can be decomposed as
\begin{equation} \label{hd}
	dX = 2\pi R\,(l_r + m_r\,\tau)\,d\hat{X}^r + dA,
\end{equation}
where $\{d\hat{X}^r\}$ form a basis of normalized harmonic 1-forms on $\Sigma$ satisfying
\begin{equation}
	\oint_{C_s} d\hat{X}^r = \delta^r{}_s.
\end{equation}
The exact term $dA = dA^1 + i\,dA^2$ transforms as a symplectic connection under area--preserving diffeomorphisms (APDs) connected to the identity. It will be useful to write the explicit form of the harmonic contributions to these 1-forms:
\begin{eqnarray}
	dX_h^s = S^s{}_u \mathbb{W}^u_r\,d\hat{X}^r,
\end{eqnarray}
where $S$ is the matrix
\begin{eqnarray}
	S:=2\pi R\left(\begin{array}{cc}
		1 & \operatorname{Re}(\tau) \\
		0 & \operatorname{Im}(\tau) 
	\end{array}\right),
\end{eqnarray}
and
\begin{eqnarray}
	\det(S\mathbb{W}) = (2\pi R)^2\operatorname{Im}(\tau) n.
\end{eqnarray}

In the original formulation \cite{Ovalle1} the $\sqrt{W}$ is defined as the pullback of the symplectic form of the target torus,
\begin{eqnarray}
	\frac{1}{2}\epsilon_{ij}\, d\hat{X}^i \wedge d\hat{X}^j
	&=& \frac{1}{2}\epsilon_{ij}\,\epsilon^{ab}\,
	\partial_a \hat{X}^i\,\partial_b \hat{X}^j\,
	d\sigma^1 \wedge d\sigma^2 = \sqrt{W}\, d\sigma^1 \wedge d\sigma^2,
\end{eqnarray}
where $a,b=1,2$. Since the central charge condition (\ref{CH}) guarantees $\sqrt{W} \neq 0$, one can consistently define the symplectic Lie bracket as 
\begin{equation}
	\{A,B\} = \frac{\epsilon^{ab}}{\sqrt{W}}\,\partial_a A\,\partial_b B.
\end{equation}

Let us recall the role of the residual gauge symmetry in this formulation. After fixing the light cone gauge, the remaining gauge transformations are the area-preserving diffeomorphisms of $\Sigma$ with respect to the volume form $\sqrt{W}\,d\sigma^1\wedge d\sigma^2$. Locally, these transformations are generated by functions on $\Sigma$ modulo constants and act on any worldvolume scalar through the Lie bracket defined above. On a compact Riemann surface, however, the APD algebra also contains a finite-dimensional harmonic sector associated with the nontrivial cohomology of $\Sigma$. When the target space is compactified, harmonic maps generally appear as dynamical fields. At the regularized level, the structure constants associated with the harmonic sector do not satisfy the Jacobi identity, obstructing its incorporation into a closed $SU(N)$ gauge algebra. By contrast, when the central charge condition is imposed, this harmonic contribution is fixed by the winding data and encoded in the harmonic part of the compact embedding. Thus, once the topological sector has been fixed, the local residual gauge symmetry acting on the dynamical fluctuations is the Hamiltonian sector of the APDs, while the harmonic sector remains as global topological data.

The Hamiltonian of the theory in $M_9\times T^2$ can be written as 
\begin{eqnarray}
	H &=& \frac{1}{2P^0_-}\int_{\Sigma}d^2\sigma\sqrt{W}\left[\left(\frac{P_m}{\sqrt{W}}\right)^2+\left(\frac{P_r}{\sqrt{W}}\right)^2  +\frac{T_{M2}^2}{2}\{X^m,X^n\}^2 + T_{M2}^2\{X^m,X^r\}^2\right. \nonumber \\ &+& \left. \frac{T_{M2}^2}{2}\{X^r,X^s\}^2-2T_{M2}P_-^0\bar{\theta}\Gamma^-\Gamma_m\{X^m,\theta\}- 2T_{M2}P_-^0\bar{\theta}\Gamma^-\Gamma_r\{X^r,\theta\}\right] \nonumber \\ 
\end{eqnarray}
subject to 
\begin{align}
	\Phi :=  d\!\left(\frac{P_m}{\sqrt{W}}\,dX^m +\frac{P_r}{\sqrt{W}}\,dX^r+ P_-^0\bar{\theta}\Gamma^- d\theta\right) &= 0, \\
	\Psi_r :=\oint_{\mathcal{C}_r}\!\left(\frac{P_m}{\sqrt{W}}\,dX^m +\frac{P_r}{\sqrt{W}}\,dX^r + P_-^0\bar{\theta}\Gamma^- d\theta\right) &= 0.
\end{align}

There exists two symmetries present in the formulation  (see \cite{mpgm3} and \cite{mpgm18} for a rigorous treatment). On the one hand,
the volume 2-form is invariant under the transformation
\begin{eqnarray}
	\hat{X}^r \rightarrow L^r_s\hat{X}^s, \quad L\in SL(2,\mathbb{Z}).
\end{eqnarray}
Under this transformation, the winding matrix $\mathbb{W}$ must transform as
\begin{eqnarray}
	\mathbb{W} \rightarrow \mathbb{W} L.
\end{eqnarray}

The second symmetry of this formulation of the supermembrane is the modular
transformation of the target torus,
\begin{eqnarray}
	\tau \rightarrow \frac{a \tau +b}{c \tau +d}, \quad
	R \rightarrow R|c\tau +d|, \quad
	A\rightarrow A\, e^{i\psi},
\end{eqnarray}
where $c \tau +d=|c\tau +d|e^{-i\psi}$ and
\begin{eqnarray}
	Q = \left(\begin{array}{cc}
		a & b \\
		c & d
	\end{array} \right) \in SL(2,\mathbb{Z}).
\end{eqnarray}
This transformation must be accompanied by a rotation of the gamma matrices
associated with the compact sector, namely
\begin{eqnarray}
	\Gamma :=\Gamma_1 + i\Gamma_2 \rightarrow \Gamma e^{i\psi}.
\end{eqnarray}
The winding matrix transforms accordingly as
\begin{eqnarray}
	\mathbb{W} \rightarrow DQD\,\mathbb{W},
\end{eqnarray}
with
\begin{eqnarray}
	D := \left(\begin{array}{cc}
		1 &  0\\
		0 & -1
	\end{array} \right).
\end{eqnarray}
Notice that $DQD\in SL(2,\mathbb{Z})$.

These two transformations show that there are two $SL(2,\mathbb{Z})$ actions
on the wrapping matrix. The one associated with the change of basis of
one--cycles on the base Riemann surface acts on the right, whereas the one
associated with the modular transformations of the target torus acts on the
left. Therefore, due to the symmetries of this formulation, and without loss of
generality, we can always find a transformation such that
\begin{eqnarray}
	\mathbb{W} \rightarrow \tilde{\mathbb{W}}
	=
	\left(\begin{array}{cc}
		l_1  & 0 \\
		0 &  l_2
	\end{array}\right).
\end{eqnarray}
Using this basis for the winding matrix, we can write
\begin{eqnarray}
	dX^1_h &=& 2\pi R
	\left(
	l_1\,d\hat{X}^1
	+
	l_2\,\operatorname{Re}(\tau)\,d\hat{X}^2
	\right),
	\\
	dX^2_h &=& 2\pi R\,l_2\,\operatorname{Im}(\tau)\,d\hat{X}^2 .
\end{eqnarray}
In what follows, we shall use this representative of the winding matrix. To
avoid overloading the notation, we omit the tildes and keep the same notation
for the transformed quantities.

After imposing the central charge condition (\ref{CH}) and the Hodge decomposition (\ref{hd}) it can be written as
\begin{eqnarray}
	H &=& \frac{8 T_{M2} \pi^4 R^4(\operatorname{Im}(\tau))^2n^2}{P_-^0} +\frac{1}{2P^0_-}\int_{\Sigma}d^2\sigma\sqrt{W}\left[\left(\frac{P_m}{\sqrt{W}}\right)^2+\left(\frac{P_r}{\sqrt{W}}\right)^2 \right. \nonumber  \\ & +& \frac{T_{M2}^2}{2}\{X^m,X^n\}^2 +\frac{T_{M2}^2}{2}\{A^r,A^s\}^2  +T_{M2}^2\left((D^rX^m)^2+(D^rA^s-D^sA^r)^2\right.   \nonumber  \\ & +& \left.2\{A^s,X_m\}D^sX^m+ \{A^r,X^m\}^2\right)+2T_{M2}^2\{A_r,A_s\}D^rA^s \nonumber  \\ & -& 2 T_{M2}P_-^0\bar{\theta}\Gamma^-\left[\Gamma_m\{X^m,\theta\}+\Gamma_r\left(D^r\theta+\{A^r,\theta\}\right)\right] \Bigg] . \nonumber \\
\end{eqnarray}
where 
\begin{eqnarray}
	D^r \bullet:= S^r_u \mathbb{W}^u_t \hat{D}^t\bullet \, , \quad \hat{D}^t\bullet := \{\hat{X}^t,\bullet\} .
\end{eqnarray}
The operator $D^r$ represents a rotation of the covariant derivative $\hat{D}^r$ (with respect to the metric $W_{rs}$) induced by the harmonic background selected by the central charge condition. This point is important because, after imposing the irreducible wrapping condition, the harmonic contribution to the compact embedding is fixed by the topology and does not describe an additional dynamical degree of freedom. The dynamical fields are instead the exact fluctuations $A^r$, together with the transverse coordinates $X^m$ and the fermionic variables.

It is also important to recall that the central charge condition affects the supersymmetry preserved by the theory. In previous analyses \cite{Restuccia7} (see also \cite{mpgm10}) it was shown that the irreducible wrapping condition implies a partial breaking of supersymmetry, since the ground-state configuration
\begin{equation}
	X^m=0, \quad X^r = X_h^r, \quad \theta=0, 
\end{equation}
must be preserved by the supersymmetry transformations. This requirement imposes a projection on the supersymmetry parameter and reduces the number of independent supersymmetries by one half. Hence, the formulation with central charge corresponds to a half-supersymmetric sector, or equivalently to an $\mathcal{N}=1$ model.

The matrix model regularization of this theory was originally obtained in \cite{mpgm8.5}. In what follows, we review this construction and then extend the formalism in order to include a proper treatment of the global APD constraint (an aspect, that, to our knowledge, has not been previously considered in the literature). This distinction is relevant because the matrix regularization captures the Hamiltonian, local sector of the area--preserving diffeomorphisms, whereas the harmonic sector keeps track of the non-trivial topology of the compact directions. In particular, the global APD constraint arises from requiring $dX^-$ to be an exact 1-form along the non-trivial one--cycles associated with the compact sector of the theory.

At this point following the procedure presented in \cite{mpgm8.5} we can express the fields in terms of a complete orthonormal basis of scalars defined over $\Sigma$, namely
\begin{eqnarray}
	X^m &=& \sum_{A}X^{m\ A }(t) Y_A, \quad A^r = \sum_{A}A^{r\ A}(t) Y_A,  \nonumber \\
	\frac{P_m}{\sqrt{W}} &=& \sum_{A}\rho_{m} ^A (t) Y_A, \quad  \frac{P_r}{\sqrt{W}} = \sum_{A}\rho_{r} ^A (t) Y_A, \nonumber \\
\end{eqnarray}
and for the fermions,
\begin{eqnarray}
	\theta = \sum_A \theta^A(t)   Y_A,
\end{eqnarray}
where $A=(a_1,a_2)$ and the case $a_1=a_2 =0$, with $Y_0 =1$, corresponds to the zero mode.  

The basis satisfies
\begin{eqnarray}
	\int_\Sigma d^2\sigma \sqrt{W} Y_AY^B = \delta_{A}^B, \quad \{Y_A,Y_B\} = g^C_{AB}Y_C ,
\end{eqnarray}
with structure constants
\begin{eqnarray}
	g^{C}_{AB} = \int_{\Sigma}d^2\sigma \sqrt{W} \{Y_A,Y_B\}Y^C,
\end{eqnarray}
and duals
\begin{eqnarray}
	Y^A := (Y_A)^* = \eta^{AB}Y_B,
\end{eqnarray}
where the matrix $\eta^{AB}$ satisfies $\eta^{AB}\eta_{BC}=\delta^A_C$ and $\eta_{AB}:=(\eta^{AB})^*$.

Now, for the compact sector, following \cite{mpgm8.5}, we then have
\begin{eqnarray}
	\hat{D}^r Y_A = \lambda^{r\ C}_A Y_C,
\end{eqnarray}
where

\begin{align}
\lambda^{r\ C}_A= \frac{1}{2\pi}\int_{\Sigma}d^2\sigma \sqrt{W}\{Y_{V^r},Y_A\}Y^{C+V^r},
\end{align}
are the coefficients that describe  the action of the compact background covariant derivatives on the APD basis. They encode the coupling between the harmonic sector fixed by the central charge and the Hamiltonian APD modes expanded in the basis 
$\{Y_A\}$,
where,
	\begin{eqnarray}
		V^r := \left\lbrace \begin{array}{cc}
			(1,0) & r=1  \\
			(0,1) & r=2 
		\end{array} \right. .
	\end{eqnarray}
The choice of the vectors $V^r$ provides a convenient way of representing the fixed harmonic contributions selected by the central charge condition.

This is precisely where the central charge formulation differs from the case analyzed in \cite{deWit4}, as originally found in \cite{mpgm8.5}. Once the irreducible winding condition has been imposed, the harmonic 1-forms are fixed by the chosen topological sector and are not expanded as dynamical fields. Therefore, the infinite-dimensional APD expansion is performed only for the dynamical fields $X^m$, $A^r$, their conjugate momenta, and the fermions. This is the reason why the $SU(N)$ matrix model description remains well-defined even in the presence of compact target-space directions.

Moreover, we assume $Y_A$ is an eigenfunction of $D_rD^r$,
\begin{eqnarray}
	\hat{D}_r\hat{D}^r Y_A = \omega_A Y_A,
\end{eqnarray}
with $\hat{D}_r = \delta_{rs} \hat{D}^s$, and no summation over $A$. This leads to
\begin{eqnarray}
	\omega_A = \delta_{rs}\lambda^{s\ C }_A\lambda^{r\ B }_C \delta_B^A ,
\end{eqnarray}
where, as before, there is no sum over the $A$ index.

Under these considerations the Hamiltonian can be re-expressed in terms of the infinite-dimensional APD structure constants as
\begin{eqnarray}
	H &=&  \frac{8 T_{M2} \pi^4 R^4(\operatorname{Im}(\tau))^2n^2}{P_-^0} + \frac{1}{2P_-^0}\Bigg[\rho^{m\ A}\rho_{m\ A}+\rho^{r\ A}\rho_{r\ A} \nonumber \\ &+& \frac{T_{M2}^2}{2}(X^{m\ A}X^{n\ B}X_{m}^CX_{n}^D + A^{r\ A}A^{s\ B}A_{r}^CA_{s}^D+X^{m\ A}A^{r\ A}X^{C}_mA^{D}_r) g^{E}_{AB}g_{CDE} \nonumber \\
	&+& T_{M2}^2(X^{m\ A}X_{m}^B\delta_{r\hat{r}}+ A^{s\ A}A^{\hat{s}\ B}\epsilon_{rs}\epsilon_{\hat{r}\hat{s}})S^r_uS^{\hat{r}}_v\mathbb{W}^u_t\mathbb{W}^v_{\hat{t}}\lambda^{t\ C}_{A} \lambda^{\hat{t}}_{BC} \nonumber \\
	&+&2T^2_{M2}(X^{m\ A}X^{B}_{m}+A^{A}_sA^{s\ B})A^{ C}_rS^r_u \mathbb{W}^u_t\lambda^{t \ D}_{A}g_{CBD} \nonumber \\
	&-&2T_{M2}P_-^0\bar{\theta}_A\Gamma^-(\Gamma_mX^{m\ B}+\Gamma_rA^{r\ B})\theta^Cg^A_{BC} \nonumber \\
	&-&2T_{M2}P_-^0\bar{\theta}_A\Gamma^-\Gamma_r\theta^B S^r_u \mathbb{W}^u_t \lambda^{t\ A}_B\Bigg].
\end{eqnarray}

The local APD constraint becomes 
\begin{eqnarray} 
	\Phi^C &=&(\rho_m^AX^{m\ B}+\rho_r^AA^{r\ B})g^{C}_{AB}-\rho_r^AS^r_u \mathbb{W}^u_t \lambda^{t\ C}_A +P_-^0\bar{\theta}_A\Gamma^-\theta^B g^{C\ A}_{\ \ \ \ \ B} =0.
\end{eqnarray}

For the global constraints at exact level, it is useful to rewrite them in the equivalent form 
\begin{eqnarray}
	\Psi_r = \int_{\Sigma}d^2\sigma\sqrt{W}\epsilon_{rs}\lbrace \hat{X}^s,X^-\rbrace =0,
\end{eqnarray}
which leads to 
\begin{eqnarray}
	\Psi_r &=& -\rho^{0}_t S^{t}_u \mathbb{W}_r^u + \epsilon_{rs}(\rho_{m\ A}X^{m\ B}+\rho_{t\ A}A^{t\ B} + P_-^0 \bar{\theta}_A\Gamma^-\theta^B )\lambda^{s \ A}_{ B} =0.
\end{eqnarray}
The sum of the zero mode of the compact momentum the associated to the center of mass of the momentum along the compact dimemsions becomes defined in terms of the rest of the fields of the compact and non-compact sectors. An important aspect of the infinite-dimensional APD rewriting is that the choice of basis depends on the topology of the spatial worldvolume. In particular, APD algebras associated with different topologies are not, in general, equivalent. In the present work we take the spatial part of the worldvolume to be a two--torus. This allows us to use a Fourier-type basis adapted to the harmonic background determined by the central charge condition.

We now use the explicit basis to simplify these expressions. We choose
\begin{eqnarray}
	Y_A := \exp{\left(2\pi ia_r \hat{X}^r\right)},
\end{eqnarray}
where $a_r$, with $r=1,2$ represent a pair of integers. This basis satisfies
\begin{eqnarray}
	Y^A = (Y_A)^* = Y_{-A}.
\end{eqnarray}
Therefore, the structure constants for a general two torus, take the form
\begin{eqnarray}
	g^C_{AB} = (2\pi i)^2 (A\times B) \delta_{A+B-C},
\end{eqnarray}
where for $A=(a_1,a_2)$ and $B=(b_1,b_2)$, 
\begin{eqnarray}
	A\times B := \epsilon^{rs}a_rb_s.
\end{eqnarray}
Furthermore, we obtain
\begin{eqnarray}
	\lambda^{r\ C}_{A} = 2\pi i (V^r \times A)  \delta_{A-C},
\end{eqnarray}
These modes should not be regarded as additional dynamical fields to be expanded in the APD basis. Rather, they encode the global winding data of the compact sector and determine how the background covariant derivatives act on the Hamiltonian APD modes.


\section{Matrix regularization of the supermembrane}
Up to this point, we have only rewritten the Hamiltonian by expanding the dynamical maps in terms of a complete orthonormal basis of functions defined on $\Sigma$. This rewriting makes explicit the infinite-dimensional APD structure of the theory, but it is still a formal expression: the sums run over infinitely many modes and no finite-dimensional regularization has yet been introduced. We now proceed to the matrix regularization of the supermembrane with central charge.

The starting point follows the construction of \cite{deWit2,deWit3,mpgm8.5}: one introduces a cut-off $\Lambda$ by restricting the APD modes to a finite subset. The aim is to approximate, at finite $\Lambda$, the Hamiltonian sector of the area--preserving diffeomorphisms, namely the sector generated by functions on $\Sigma$ modulo constants.

This qualification is important in the central charge case. On a compact Riemann surface, the residual APDs contain both a Hamiltonian sector and a finite-dimensional harmonic sector. The former is the sector represented by the Poisson algebra of the basis functions $Y_A$ and is the one that admits the usual matrix approximation. The latter is fixed by the winding data and is encoded in the harmonic background $dX_h^r$; it is not regularized as an independent local gauge degree of freedom.

Thus, denoting by $f_{ABC}$ the structure constants of the finite matrix algebra, the regularization is chosen so that
\begin{eqnarray}
	\lim_{\Lambda \rightarrow \infty} f_{ABC} = g_{ABC},
\end{eqnarray}
for the Hamiltonian APD structure constants $g_{ABC}$.

For the regularized supermembrane without windings, it was shown that the structure group of the theory corresponds to 
\begin{eqnarray}
	G_{\Lambda} = SU(N), \qquad \Lambda = N^2 - 1.
\end{eqnarray}
When compact directions are taken into account, the authors of \cite{deWit4} were unable to find an $SU(N)$ extension satisfying the Jacobi identity. However, in the particular case of the supermembrane with central charge, it was shown in \cite{mpgm8.5} that such an extension is not needed, and that there exists a representation of the $SU(N)$ generators allowing for a well-defined matrix regularization. The key difference between the results presented in \cite{deWit4} and those in \cite{mpgm8.5} lies in the treatment of the harmonic contribution to the compact fields. In what follows, we shall summarize the results of \cite{mpgm8.5}.

The first step is to write the generators $T_A$ as
\begin{eqnarray}
	T_{0} &=& c\, N\, \mathbb{I}, \\
	T_{A} &=& c\, N\, \omega^{\frac{1}{2}a_1 a_2} P^{a_1} Q^{a_2}, \quad A \neq 0,
\end{eqnarray}
where $P$ and $Q$ are the Heisenberg matrices satisfying
\begin{eqnarray}
	PQ = \omega QP, \qquad \omega = e^{\frac{2\pi i}{N}}.
\end{eqnarray}
The generators satisfy
\begin{eqnarray}
	T_{A} T_{B} &=& c\, N\, \omega^{\frac{1}{2} A \times B} T_{A+B}, \\
	\mathrm{Tr}(T_A) &=& c\, N^2 \delta_A, \\
	\mathrm{Tr}(T_A T_B) &=& c^2 N^3 \delta_{A+B},
\end{eqnarray}
and
\begin{eqnarray}
	[T_A, T_B] = f^{A+B}_{AB} T_{A+B},
\end{eqnarray}
with

In addition, the structure constants with one index associated with the compact space are defined by
\begin{eqnarray}
	[T_{V^r}, T_A] = f^{V^r + A}_{V^r A} T_{V_r + A}.
\end{eqnarray}
This decomposition provides a $SU(N)$ regularization of the compactified M2-brane with central charge.
Notice that the limit $N \rightarrow \infty$ cannot be taken directly in order to relate $f_{ABC}$ with $g_{ABC}$. Instead, in order to have a well-defined limit, one must fix an energy level $E$ and then take the limit $N \rightarrow \infty$ while keeping $E$ fixed. In practice, this corresponds to considering the limit $N \rightarrow \infty$ while keeping $(a_1, a_2)$ fixed. Under these assumptions, it is straightforward to check that
\begin{eqnarray}
	\lim_{N \rightarrow \infty} f^C_{AB} = -2 i \pi c (A \times B) \delta_{A+B-C}.
\end{eqnarray}
Therefore, one can fix
\begin{eqnarray}
	c = -2\pi i
\end{eqnarray}
in order to obtain full correspondence with $g^C_{AB}$. 

Notice that this already ensures the convergence of the structure constants associated with the compact sector,
\begin{eqnarray}
	\lim_{N\rightarrow \infty}
	f^{V^r + C}_{V^r A}
	=
	(2 \pi i)^2 (V^r \times A) \delta_{A-C}
	=
	g^{V^r + C}_{V^r A}
	=
	2\pi i \lambda^{rC}_A.
\end{eqnarray}

Finally we can write the regularized Hamiltonian as
\begin{eqnarray}
	H &=&  \frac{8 T_{M2} \pi^4 R^4(\operatorname{Im}(\tau))^2n^2}{P_-^0} + \frac{1}{2P_-^0}\Bigg[\rho^{m\ A}\rho_{m\ A}+\rho^{r\ A}\rho_{r\ A} \nonumber \\ &+& \frac{T_{M2}^2}{2}(X^{m\ A}X^{n\ B}X_{m}^CX_{n}^D + A^{r\ A}A^{s\ B}A_{r}^CA_{s}^D) f^{E}_{AB}f_{CDE} \nonumber \\
	&+& \frac{T_{M2}^2}{(2\pi i)^2}(X^{m\ A}X_{m}^B\delta_{r\hat{r}} +A^{s\ A}A^{\hat{s}\ B}\epsilon_{rs}\epsilon_{\hat{r}\hat{s}})S^r_uS^{\hat{r}}_v\mathbb{W}^u_t\mathbb{W}^v_{\hat{t}}f^{V^t + C}_{V^t A} f^{V^{\hat{t}} -C}_{V^{\hat{t}} B} \nonumber \\
	&+&\frac{T^2_{M2}}{\pi i}(X^{m\ A}X^{B}_{m}+A^{A}_sA^{s\ B})A^{ C}_rS^r_u \mathbb{W}^u_tf^{V^t + D}_{V^t A}f_{CBD} \nonumber \\
	&-&2T_{M2}P_-^0\bar{\theta}_A\Gamma^-(\Gamma_mX^{m\ B}+\Gamma_rA^{r\ B})\theta^Cf^A_{BC} \nonumber \\
	&-&\frac{T_{M2}}{\pi i}P_-^0\bar{\theta}_A\Gamma^-\Gamma_r\theta^B S^r_u \mathbb{W}^u_t f^{V^t + A}_{V^t B}\Bigg].
\end{eqnarray}

together with the local constraint
\begin{eqnarray}
	\Phi^C&=& (\rho_m^AX^{m\ B}+\rho_r^AA^{r\ B})f^{C}_{AB}-\frac{1}{2\pi i}\rho_r^AS^r_u \mathbb{W}^u_t f^{V^t+ C}_{V^t \ A} +P_-^0\bar{\theta}_A\Gamma^-\theta^B f^{C\ A}_{\ \ \ \ \ B}  =0,
\end{eqnarray}
and the global one,
\begin{eqnarray}
	\Psi_r &= & -\rho^{0}_t S^{t}_u \mathbb{W}_r^u + \frac{\epsilon_{rs}}{2\pi i}(\rho_{m\ A}X^{m\ B}+\rho_{t\ A}A^{t\ B} + P_-^0 \bar{\theta}_A\Gamma^-\theta^B )f^{V^s+ A}_{V^s \ B}=0. 
\end{eqnarray}
Here we can see that the global constraint can be solved in order to determine the zero modes of the momenta associated with the compact directions. Explicitly, we obtain
\begin{eqnarray}
	\rho^{0}_t &=& \frac{\epsilon_{rs}}{2\pi i}f^{V^s+ A}_{V^s \ B}(\mathbb{W}^{-1}S^{-1})^r_t(\rho_{m\ A}X^{m\ B}+\rho_{v\ A}A^{v\ B} + P_-^0 \bar{\theta}_A\Gamma^-\theta^B ).
\end{eqnarray}

One of the most important questions concerning the regularized Hamiltonian is whether its spectrum is discrete or continuous. This issue is particularly relevant since the supersymmetric regularized Hamiltonian of the supermembrane in flat eleven-dimensional space is known to have a continuous spectrum \cite{dwln}. Remarkably, the presence of a central charge drastically alters this behavior. Indeed, the central charge condition guarantees the existence of nonvanishing quadratic terms in the Hamiltonian, even when string-like configurations are considered. As a consequence, for the particular Hamiltonian considered here,\footnote{On general grounds, the existence of quadratic terms by themselves is not sufficient to guarantee the discreteness of the spectrum.} the spectrum of the regularized Hamiltonian with central charge is discrete and has finite multiplicity. A rigorous proof of this result can be found in \cite{mpgm11}.

This constitutes a significant improvement for the theory, as it allows the supermembrane to describe at least a sector of M-theory, which was the original motivation for its formulation. Nevertheless, a final and essential issue remains to be addressed. One must verify that, in the limit $N \rightarrow \infty$, the spectrum of the regularized Hamiltonian coincides with that of the non-regularized theory. This is a highly nontrivial problem, for which no general answer is known, even in the simplest case of the supermembrane in flat space (see \cite{dwln}). In the next section, we address this question for a simplified version of the regularized Hamiltonian with central charge. Specifically, we consider the semiclassical limit, since this is the regime in which the exact spectrum can be computed explicitly. This simplified setting provides a useful laboratory to test the large $N$ properties of the regularized model before addressing the corresponding problem for the full interacting Hamiltonian in a future work.

We close this section by summarizing the formal content of the previous
construction, emphasizing the points that are relevant for the $SU(N)$ matrix
regularization of the supermembrane with central charge.

\subsubsection*{Summary of the regularization with central charge}
Let $\Sigma$ be a compact Riemann surface of genus $g$, endowed with the volume
form $\omega=\sqrt{W}\,d\sigma^1 \wedge d\sigma^2$, and consider the light cone gauge
supermembrane where the space of embbeding maps from the base manifold to the compact sector of the target space satisfy the central charge condition. Then:

\begin{enumerate}
	\item The residual infinitesimal area--preserving diffeomorphisms split into a Hamiltonian sector and a harmonic sector. The Hamiltonian sector is generated by functions modulo constants,
	\begin{equation}
		C^\infty(\Sigma)/\mathbb{R},
	\end{equation}
	with Lie bracket given by 
	\begin{equation}
		\{F,G\}=
		\frac{\epsilon^{ab}}{\sqrt{W}}\partial_aF\partial_bG.
	\end{equation}
	The harmonic sector is finite-dimensional, of dimension $2g$, and is associated with the non-trivial cohomology of $\Sigma$.
	
	\item In the central charge sector, the compact embedding decomposes as
	\begin{equation}
		dX^r=dX_h^r+dA^r,
	\end{equation}
	where $dX_h^r$ is fixed by the winding data and a $dA^r$ that contains the
	dynamical fluctuations.Thus, the harmonic contribution carries the
	topological data responsible for the central charge, while the exact part describes the local dynamical degrees of freedom.
	\item Once the topological sector has been fixed, the local residual gauge symmetry acting on the dynamical fields is the Hamiltonian APD sector. This is the sector that is regularized by matrices. In particular, after expanding the dynamical fields in a basis $Y_A$, the regularization replaces
	\begin{equation}
		Y_A \longrightarrow T_A,
	\end{equation}
	and approximates the Lie algebra by the matrix commutator algebra,
	\begin{equation}
		\{Y_A,Y_B\}=g^C{}_{AB}Y_C
		\quad \longrightarrow \quad
		[T_A,T_B]=f^C{}_{AB}T_C,
	\end{equation}
	with $f^C{}_{AB}\rightarrow g^C{}_{AB}$ in the large $N$ limit, after the appropriate normalization and for a fixed energy level. This includes those associated with the compact sector. 
	\begin{equation}
		\{Y_{V^r},Y_A\}=g^{C + V^r}_{V^r A}Y_C
		\quad \longrightarrow \quad
		[T_{V_r},T_A]=f^{C+V^r}_{{V_r}A}T_C.
	\end{equation}
	
	\item The matrix model therefore regularizes the Hamiltonian APD algebra over a fixed non-trivial harmonic background. The harmonic sector itself is not regularized as an independent local gauge degree of freedom; instead, it enters through the background $dX_h^r$, the covariant derivatives
	\begin{equation}
		D^r\bullet=\{X_h^r,\bullet\},
	\end{equation}
	and there are found a  distinguished $SU(N)$ compact generators $T_{V^r}$ associated with the winding directions.
	
	\item The global APD constraint remains as the condition associated with the zero periods of the closed 1-form $dX^-$ along the non-trivial one cycles of $\Sigma$. In the regularized theory, this constraint determines the zero modes of the momenta conjugated in the compact directions, that become restricted by the contributions of the other dynamical fields of the theory. Hence, the large $N$ limit recovers the Hamiltonian APD dynamics, while the harmonic sector and the global constraint retain the topological information fixed before the regularization.
\end{enumerate}

\section{Semiclasical limit of the supermembrane Hamiltonian}
In this section, we present new results concerning the regularized and non-regularized supersymmetric semiclassical Hamiltonians of the supermembrane. Before starting the analysis, let us specify the procedure that will be followed. The idea is to consider a classical solution of the supermembrane equations of motion and to expand all fields around this solution, namely,
\begin{eqnarray}
	X^{\tilde{m}} = X^{\tilde{m}}_{\mathrm{cl}} + \tilde{X}^{\tilde{m}}, \qquad 
	P_{\tilde{m}} = P^{\mathrm{cl}}_{{\tilde{m}}} + \tilde{P}_{\tilde{m}},
\end{eqnarray}
and similarly for the fermionic variables,
\begin{eqnarray}
	\theta = \theta_{\mathrm{cl}} + \Psi.
\end{eqnarray}
Our choice of classical solution is given by
\begin{eqnarray}
	X^m_{\mathrm{cl}} = P^{\mathrm{cl}}_m = P^{\mathrm{cl}}_r = \theta_{\mathrm{cl}} = 0,
\end{eqnarray}
and 
\begin{eqnarray}
	X^r_{\mathrm{cl}} = X^r_h = (S\mathbb{W})^r_{\ s}\,\hat{X}^s.
\end{eqnarray}
It is a straightforward computation to verify that this configuration indeed solves the supermembrane equations of motion.

Now, expanding the Hamiltonian around the classical solution, we obtain the following expansion:
\begin{eqnarray}
	H = H_{0} + H_{1} + H_{2} + \cdots,
\end{eqnarray}
together with the corresponding expansion of the constraints,
\begin{eqnarray}
	\Phi &=& \Phi_0 + \Phi_1 + \Phi_2+ \cdots, \\
	\Psi_r &=& \Psi_r^{\,0} + \Psi_r^{\,1} + \Psi_r^{\,2}+ \cdots.
\end{eqnarray}
Here, the subscript (or superscript, in the case of $\Psi_r$) denotes the order of the terms in the expansion.

Now, the zeroth-order contributions to the constraints vanish, since the classical solution satisfies the constraints. On the other hand, the first-order contribution to the Hamiltonian can be written as
\begin{eqnarray}
	H_{1} = 
	\left.\frac{\partial H}{\partial X^{\tilde{m}}}\right|_{\mathrm{cl\,sol}} \tilde{X}^{\tilde{m}}
	+ \left.\frac{\partial H}{\partial P_{\tilde{m}}}\right|_{\mathrm{cl\,sol}} \tilde{P}_{\tilde{m}}
	+ \left.\frac{\partial H}{\partial \theta}\right|_{\mathrm{cl\,sol}} \bar\theta,
\end{eqnarray}
which vanishes identically in our case. Therefore, as expected, the semiclassical limit of the Hamiltonian is obtained by retaining only the zeroth- and quadratic-order terms in the Hamiltonian, together with the linear terms in the constraints.


\subsection{Semiclassical limit of supermembrane Hamiltonian}
In this subsection, we analyze the semiclassical limit of the non-regularized supersymmetric Hamiltonian. Previously, in \cite{mpgm17}, the corresponding bosonic Hamiltonian had been analyzed. In the present work we extend those results to the supersymmetric case. This extension is relevant because the fermionic sector cancels the usual bosonic zero-point contribution mode by mode, removing the divergent constant associated with the infinite sum of oscillator modes which appears in the purely bosonic description. Our goal is to extract the effective Hamiltonian governing the quadratic fluctuations around a classical background configuration that satisfies the equations of motion and the constraints. By explicitly solving the constraints at this order and identifying the appropriate canonical variables, we obtain a semiclassical Hamiltonian that will serve as a reference for the comparison with the regularized theory in the large-$N$ limit.

Following the procedure presented at the beginning of this section, we find that the semiclassical limit of the non-regularized Hamiltonian is given by
\begin{eqnarray}
	H^{\mathrm{scl}} &=&  \frac{8 T_{M2} \pi^4 R^4(\operatorname{Im}(\tau))^2n^2}{P_-^0} + \frac{1}{2P_-^0}\Bigg[\rho^{m\ A}\rho_{m\ A}+\rho^{r\ A}\rho_{r\ A}  + T_{M2}^2(X^{m\ A}X_{m}^B\delta_{r\hat{r}} \nonumber \\ &+& A^{s\ A}A^{\hat{s}\ B}\epsilon_{rs}\epsilon_{\hat{r}\hat{s}})S^r_uS^{\hat{r}}_v\mathbb{W}^u_t\mathbb{W}^v_{\hat{t}}\lambda^{t\ C}_{A} \lambda^{\hat{t}}_{BC} -2T_{M2}P_-^0\bar{\theta}_A\Gamma^-\Gamma_r\theta^B S^r_u \mathbb{W}^u_t \lambda^{t\ A}_B\Bigg].
\end{eqnarray}
\label{Hnreg}
together with the constraints
\begin{eqnarray}
	\rho_r^AS^r_u \mathbb{W}^u_t \lambda^{t\ C}_A = 0,
\end{eqnarray}
\begin{eqnarray}
	\rho^{0}_t S^{t}_u \mathbb{W}_r^u = 0.
\end{eqnarray}

From these expressions we see that the only information coming from the global constraint is that the sum of the KK zero modes $\rho^{0s}$ weighted by the wrapping numbers must vanish. Let us now analyze the local constraint, which has to be satisfied independently for each value of $C=(c_1,c_2)$. Using the explicit expressions for $S$, $\mathbb{W}$ and $V^t$, the two coefficients entering the constraint are
	\begin{eqnarray}
		(S\mathbb{W})^1_t(V^t\times C)
		&=&2\pi R\left(l_1c_2-l_2\operatorname{Re}(\tau)c_1\right),
		\\
		(S\mathbb{W})^2_t(V^t\times C)
		&=&-2\pi Rl_2\operatorname{Im}(\tau)c_1.
	\end{eqnarray}
	
	We first consider the case $\operatorname{Re}(\tau)\neq0$. In this case, for non-zero modes and assuming
	\begin{eqnarray}
		l_1c_2-l_2\operatorname{Re}(\tau)c_1\neq0,
	\end{eqnarray}
	the local constraint can be solved as
	\begin{eqnarray}
		\rho_1^C = -\frac{\rho_2^C(S\mathbb{W})^2_t(V^t\times C)}{(S\mathbb{W})^1_t(V^t\times C)},
	\end{eqnarray}
	for each value of $C$. Notice that this expression also includes the modes $C=(0,c_2)$, with $c_2\neq0$. In this case $(S\mathbb{W})^2_t(V^t\times C)=0$, and therefore the constraint simply gives
	\begin{eqnarray}
		\rho_1^{(0,c_2)}=0.
	\end{eqnarray}
	Since the semiclassical Hamiltonian is quadratic in the momenta, Hamilton's equations imply that
	\begin{eqnarray}
		\dot{A}^{rA}\propto \rho^{rA}.
	\end{eqnarray}
	Consequently, for these modes we also have
	\begin{eqnarray}
		\dot{A}^{1(0,c_2)}=0.
	\end{eqnarray}
	Therefore, these modes do not introduce an additional contribution when the kinetic term is rewritten in terms of the reduced canonical variables.
	
	Analyzing the kinetic term of the action,
	\begin{eqnarray}
		&&\int_{\Sigma} P_r \dot{A}^r
		= -  \int_{\Sigma}\dot{P}_r A^r =-\rho_{rA}\dot{A}^{rA}
		=  \sum_A\left(\frac{\dot{\rho}_{2}^A}{(S\mathbb{W})^1_t(V^t\times A)}\right) 
		(\epsilon_{rs}A^{r}_A(S\mathbb{W})^s_t(V^t\times A)),\nonumber \\ &&
	\end{eqnarray}
	we can identify a new pair of canonically conjugate variables associated to the compact extra dimensions defined as
	\begin{eqnarray}
		\hat{\rho}_{A} &:=&
		T_{M2} \epsilon_{rs}A^{r}_A(S\mathbb{W})^s_t(V^t\times A)
		, \\
		\hat{X}^{A} &:=& \frac{1}{T_{M2} }
		\left(\frac{\rho_{2}^A}{(S\mathbb{W})^1_t(V^t\times A)}\right).
	\end{eqnarray}
	
	Let us now consider the case $\operatorname{Re}(\tau)=0$. For $c_2\neq0$, the coefficient $(S\mathbb{W})^1_t(V^t\times C)$ remains non-vanishing and the previous solution of the local constraint, as well as the corresponding definition of the canonical pair, can be used without modification. The only additional non-zero modes that have to be considered are therefore $C=(c_1,0)$, with $c_1\neq0$. For these modes,
	\begin{eqnarray}
		(S\mathbb{W})^1_t(V^t\times C)=0,
		\qquad
		(S\mathbb{W})^2_t(V^t\times C)\neq0,
	\end{eqnarray}
	and the local constraint gives
	\begin{eqnarray}
		\rho_2^{(c_1,0)}=0.
	\end{eqnarray}
	Again, Hamilton's equations imply
	\begin{eqnarray}
		\dot{A}^{2(c_1,0)}=0.
	\end{eqnarray}
	For these modes the reduction can therefore be performed using the other non-vanishing component. Equivalently, one can define the corresponding canonical pair by interchanging the roles of the two compact directions. This only amounts to a constant rescaling of the reduced canonical variables and leads to the same contribution to the semiclassical Hamiltonian. Hence, when the sum over all Fourier modes is performed, the final reduced Hamiltonian has the same form in both cases $\operatorname{Re}(\tau)\neq0$ and $\operatorname{Re}(\tau)=0$.
	
	Finally, for $C=(0,0)$ both coefficients entering the local constraint vanish and the constraint reduces to the trivial identity $0=0$. This mode is instead controlled by the global constraint discussed above, which fixes the compact momentum zero mode.
	
	Then, the final Hamiltonian in the semiclassical limit can be written as
	\begin{eqnarray}
		H^{\mathrm{scl}} &=&  \frac{8 T_{M2} \pi^4 R^4(\operatorname{Im}(\tau))^2n^2}{P_-^0} + \frac{1}{2P_-^0}\Bigg[\rho^{\hat{m}A}\rho_{\hat{m}A}\nonumber \\
		&+& 4\pi^2T_{M2}^2\sum_A X^{\hat{m}A}X^{\hat{m}}_A\delta_{r\hat{r}} S^r_uS^{\hat{r}}_v\mathbb{W}^u_t\mathbb{W}^v_{\hat{t}} (V^t\times A)  (V^{\hat{t}}\times A)\nonumber \\
		&-&4 \pi iT_{M2}P_-^0\sum_A\bar{\theta}_A\Gamma^-\Gamma_r\theta^A S^r_u \mathbb{W}^u_t (V^t\times A) \Bigg].
	\end{eqnarray}
	where $X^{\hat{m}}= (X^m,\hat{X})$ and the same for the $\rho's$.

	At this point we can perform further transformations. In first place we can always do a transformation of the form $\rho\rightarrow l\rho$ and $X\rightarrow Xl^{-1}$ since this will not change the Poisson bracket structure. We can use this transformation with $l=\sqrt{2P_-^0}$ to eliminate that coeficcient from the kinetic term.
	
	On the other hand, in order to identify more clearly the structure of the
	fermionic contributions to the semiclassical Hamiltonian, we define
	\begin{eqnarray}
		\hat{\Gamma}_A : = \frac{\Gamma_rS^r_u \mathbb{W}^u_t (V^t\times A)}{\sqrt{\delta_{r\hat{r}} S^r_uS^{\hat{r}}_v\mathbb{W}^u_t\mathbb{W}^v_{\hat{t}} (V^t\times A)  (V^{\hat{t}}\times A)}}, 
	\end{eqnarray}
	which satisfy $\hat{\Gamma}_A^2=\mathbb{I}_{32}$ and 
	
	\begin{eqnarray}
		\{\hat{\Gamma}_A,\Gamma_m \} =0.
	\end{eqnarray}
	Now, using the following representation 
	
	\begin{eqnarray*}
		\Gamma^+ &=& \sqrt{2}\left(\begin{array}{cc}
			0 & 0 \\
			\mathbb{I}_{16} & 0 
		\end{array}\right), \quad \Gamma^- = \sqrt{2}\left(\begin{array}{cc}
			0 & \mathbb{I}_{16} \\
			0 & 0 
		\end{array}\right), \quad  \Gamma^M = \left(\begin{array}{cc}
			\gamma^M  & 0 \\
			0 & -\gamma^M 
		\end{array}\right), 
	\end{eqnarray*}
	where $\gamma^M$  are the Euclidean $SO(9)$ gamma matrices, we can solve the condition $\Gamma^+\theta=0$ to get
	
	\begin{eqnarray}
		\theta_A = \Lambda \left(\begin{array}{c}
			0  \\
			\psi 
		\end{array}\right), \quad |\Lambda|^2:= \frac{1}{2\sqrt{2}P_-^{0}}.
	\end{eqnarray}
	Then we can now write the following,
	\begin{eqnarray}
		\bar{\theta}_A\Gamma^-\Gamma_rS^r_u \mathbb{W}^u_t (V^t\times A)\theta^A = - \sqrt{2}|\Lambda|^2 \omega_A\psi^\dagger_A \hat{\gamma}_A \psi^A,
	\end{eqnarray}
	with 
	
	\begin{eqnarray}
		\hat{\gamma}_A : = (\omega_A)^{-1}\gamma_rS^r_u \mathbb{W}^u_t (V^t\times A), \quad \omega_A =\sqrt{\delta_{r\hat{r}} S^r_uS^{\hat{r}}_v\mathbb{W}^u_t\mathbb{W}^v_{\hat{t}} (V^t\times A)  (V^{\hat{t}}\times A)},  
	\end{eqnarray}
	with $\hat{\gamma}_A^2 = \mathbb{I}_{16}.$ 
	Since $\hat{\gamma}_A$ is invertible we can write
	
	\begin{eqnarray}
		\hat{\gamma}_A := O^T_A u \ O_A, \quad u:=\left(\begin{array}{cc}
			\mathbb{I}_8 & 0 \\
			0&-\mathbb{I}_8 
		\end{array}\right)
	\end{eqnarray}
	and write
	\begin{eqnarray}
		\bar{\theta}_A\Gamma^-\Gamma_rS^r_u \mathbb{W}^u_t (V^t\times A)\theta^A = \chi^\dagger_A \ u \ \chi^A = \Theta^{\dagger}_{1A} \Theta_{1}^A - \Theta^{\dagger}_{2 A} \Theta_2^{ A},
	\end{eqnarray}
	with $\chi:= O \psi$, and 
	
	\begin{eqnarray}
		\chi^A = \left(\begin{array}{c}
			\Theta_{1}^A  \\
			\Theta_{2}^A 
		\end{array} \right).
	\end{eqnarray}
	Then the final Hamiltonian is given by 
	
	\begin{eqnarray}
		H^{scl} &=&
		\frac{8 T_{M2} \pi^4 R^4(\mathrm{Im}(\tau))^2 n^2}{P_-^0} + \rho^{\hat{m}A}\rho_{\hat{m}A} + \sum_A\left[\omega_A^2X^{\hat{m}A}X^{\hat{m}}_A-i\omega_A(\Theta^{\dagger}_{1A} \Theta_{1}^A - \Theta^{\dagger}_{2 A} \Theta_2^{ A})\right], \nonumber \\ &&
	\end{eqnarray}
	where 
	
	\begin{eqnarray}
		\omega_A^2
		&=&
		\frac{\pi^2}{(P_-^0)^2}T_{M2}^2 \delta_{r\hat{r}} S^r_uS^{\hat{r}}_v\mathbb{W}^u_t\mathbb{W}^v_{\hat{t}} (V^t\times A)  (V^{\hat{t}}\times A), \\
		&=& \frac{4\pi^4R^2T_{M2}^2}{(P_-^0)^2} (l^2_1a_2^2-2l_1l_2\operatorname{Re}(\tau)a_1a_2+|\tau|^2l^2_2a^2_1)
	\end{eqnarray}
	Therefore, the semiclassical Hamiltonian can be interpreted as the Hamiltonian of a supersymmetric harmonic oscillator, shifted by the constant topological contribution induced by the central charge condition. This shift is given by
	\begin{equation}
		\frac{8 T_{M2} \pi^4 R^4(\mathrm{Im}(\tau))^2 n^2}{P_-^0},
	\end{equation}
	and provides the lower energy bound associated with the irreducible wrapping. The oscillator frequencies are determined by the winding data and by the moduli of the target torus through $\omega_A$.
	
	It is important to notice that this semiclassical Hamiltonian preserves the same amount of supersymmetry as the original supermembrane with central charge. As discussed above, the central charge condition fixes the harmonic background and breaks one half of the original supersymmetries. Hence, the semiclassical Hamiltonian still describes an $\mathcal{N}=1$ supersymmetric model, now realized explicitly as a supersymmetric harmonic oscillator around the fixed topological sector.
	
	It is worth mentioning that the semiclassical spectrum of the supermembrane had already been analyzed in \cite{Duff5}, without explicitly imposing the topological restriction associated with the central charge condition and without considering the corresponding regularized model. In that case, the semiclassical Hamiltonian was also reduced to the form of a harmonic oscillator, with frequencies determined by the winding numbers and by the radii of the compact directions. In the notation of \cite{Duff5}, these frequencies are written as
	\begin{equation}
		\omega^2_{mn} = (m\,l_2 R_2)^2+(n\,l_1 R_1)^2.
	\end{equation}
	where a central charge condition was implicitly assumed although the modifications required in the complete Hamiltonian were not incorporated. Our result also generalizes  this semiclassical spectrum to the case in which the target torus is not necessarily rectangular. Indeed, the dependence on the real part of the complex modulus $\tau$ produces the additional mixing term in $\omega_A^2$. When one restricts to the rectangular case,
	\begin{equation}
		\operatorname{Re}(\tau)=0,
	\end{equation}
	the mixing term vanishes and the previous expression is recovered, up to the corresponding identifications of notation and normalization conventions.

	\subsection{Semiclassical limit of the supermembrane regularized Hamiltonian}
	
	As in the non-regularized case, the solution of the local constraint requires a separate consideration of the different Fourier modes. The same cases discussed in the previous subsection arise in the regularized theory, with the corresponding coefficients now expressed in terms of the finite-$N$ trigonometric functions. The analysis of these particular modes is completely analogous and leads to the same form of the reduced semiclassical Hamiltonian. Therefore, in order to avoid repeating the discussion, in what follows we explicitly consider the case $\operatorname{Re}(\tau)\neq0$, for which the regularized local constraint can be solved in the same way as in the generic continuum case. In this case the new conjugated pair associated with the compactified dimensions can be defined as 
	
	\begin{eqnarray}
		\hat{\rho}_{A} &:=&
		T_{M2} \epsilon_{rs}A^{r}_A(S\mathbb{W})^s_t\sin\left(\dfrac{\pi(V^t\times A)}{N}\right)
		, \\
		\hat{X}^{A} &:=& \frac{1}{T_{M2} }
		\left(\frac{\rho_{2}^A}{(S\mathbb{W})^1_t}\sin\left(\dfrac{\pi(V^t\times A)}{N}\right)\right).
	\end{eqnarray}
	together with 
	
	\begin{eqnarray}
		\hat{\Gamma}^{reg}_A
		&:=&
		\frac{\Gamma_rS^r_u \mathbb{W}^u_t \sin\left(\dfrac{\pi(V^t\times A)}{N}\right)}{\sqrt{\delta_{r\hat{r}} S^r_uS^{\hat{r}}_v\mathbb{W}^u_t\mathbb{W}^v_{\hat{t}} \sin\left(\dfrac{\pi(V^t\times A)}{N}\right)  \sin\left(\dfrac{\pi(V^{\hat{t}}\times A)}{N}\right)}}, \nonumber  \\ &&
	\end{eqnarray}
	Then we obtain 
	
	\begin{eqnarray}
		H^{scl}_{reg} &=&
		\frac{8 T_{M2} \pi^4 R^4(\mathrm{Im}(\tau))^2 n^2}{P_-^0} + \rho^{\hat{m}A}\rho_{\hat{m}A} + \sum_A\left[\omega^2_{NA}X^{\hat{m}A}X_{\hat{m}A} + i \omega_{NA}(\Theta^{\ T}_{1A} \Theta_{1}^A - \Theta_{2A}^T \Theta_2^A) \right], \nonumber \\&&
	\end{eqnarray}
	where
	\begin{eqnarray}
		\omega^2_{NA}
		&=&
		\frac{4\pi^2R^2 N^2 T_{M2}^2}{(P_-^0)^2} \Bigg(l^2_1\sin^2\left(\dfrac{\pi a_2}{N}\right) -2l_1l_2\operatorname{Re}(\tau)\sin\left(\dfrac{\pi a_1}{N}\right)\sin\left(\dfrac{\pi a_2}{N}\right) \nonumber +l_2^2|\tau|^2\sin^2\left(\dfrac{\pi a_1}{N}\right)\Bigg).
	\end{eqnarray}
	As expected, the regularized semiclassical Hamiltonian also takes the form of a supersymmetric harmonic oscillator. The effect of the matrix regularization is encoded in the replacement of the continuum frequencies $\omega_A$ by the finite-$N$ frequencies $\omega_{NA}$, where the dependence on the APD modes is now expressed through trigonometric functions. Therefore, the regularized model preserves the oscillator structure of the continuum semiclassical theory, while introducing the finite-$N$ deformation that will be relevant in the large-$N$ comparison.
	\subsection{Correspondence with the Continuum Theory in the Large-$N$ Limit}
	
	As expected, the $SU(N)$ matrix regularization reproduces the APD structure constants in the large-$N$ limit,\ cite{mpgm8.5},
	\begin{equation}
		f^C{}_{AB}\longrightarrow g^C{}_{AB},\qquad
		f^{C+V^r}_{V^rA}\longrightarrow g^{C+V^r}_{V^rA},
		\qquad N\rightarrow \infty.
	\end{equation}
	Here, the limit is understood by keeping an energy level fixed. More precisely, the Fourier labels $A=(a_1,a_2)$ associated with the APD basis are kept fixed while $N\rightarrow\infty$. With this prescription, the finite-dimensional matrix algebra approximates the continuum APD algebra. The remaining question is whether this algebraic correspondence is also reflected at the level of the spectrum.
	
	In the full interacting theory, this is a highly non-trivial problem. Here, we address it in the semiclassical approximation discussed above. In this regime, both the non-regularized and regularized Hamiltonians reduce to supersymmetric harmonic oscillators, with frequencies $\omega_A$ and $\omega_{NA}$, respectively. Therefore, for each fixed mode $A$, the eigenvalues take the general form
	\begin{equation}
		E_A =
		E_{\mathrm{top}}
		+
		\omega_A
		\left(
		n_A + N_{F,A}
		\right),
	\end{equation}
	for the continuum theory, and
	\begin{equation}
		E_{NA} =
		E_{\mathrm{top}}
		+
		\omega_{NA}
		\left(
		n_A + N_{F,A}
		\right),
	\end{equation}
	for the regularized theory. Here, $E_{\mathrm{top}}$ denotes the constant topological contribution induced by the central charge condition, $n_A$ is the bosonic occupation number, and $N_{F,A}$ denotes the corresponding fermionic occupation number. Hence, in the semiclassical approximation, the comparison between the two spectra reduces to the comparison between the corresponding frequencies in the large-$N$ limit.
	
	For the non-regularized Hamiltonian, the frequency is
	\begin{eqnarray}
		\omega_A^2
		&=&
		\frac{4\pi^4R^2T_{M2}^2}{(P_-^0)^2}
		\Big(
		l^2_1a_2^2
		+l^2_2\operatorname{Im}(\tau)^2a^2_1
		-2l_1l_2\operatorname{Re}(\tau)a_1a_2
		+l_2^2\operatorname{Re}(\tau)^2a_1^2
		\Big),
	\end{eqnarray}
	whereas in the regularized case one obtains
	\begin{eqnarray}
		\omega^2_{NA}
		&=&
		\frac{4\pi^2R^2 N^2 T_{M2}^2}{(P_-^0)^2}
		\Bigg[
		l^2_1
		\sin^2\!\left(\dfrac{\pi a_2}{N}\right)
		+
		l^2_2\operatorname{Im}(\tau)^2
		\sin^2\!\left(\dfrac{\pi a_1}{N}\right)
		\nonumber \\
		&&
		-2l_1l_2\operatorname{Re}(\tau)
		\sin\!\left(\dfrac{\pi a_1}{N}\right)
		\sin\!\left(\dfrac{\pi a_2}{N}\right)
		\nonumber \\
		&&
		+
		l_2^2\operatorname{Re}(\tau)^2
		\sin^2\!\left(\dfrac{\pi a_1}{N}\right)
		\Bigg].
	\end{eqnarray}
	Thus, the semiclassical spectral convergence is controlled by the large-$N$ behavior of $\omega^2_{NA}$. Using
	\begin{equation}
		N\sin\!\left(\frac{\pi a_r}{N}\right)
		\longrightarrow
		\pi a_r,
		\qquad N\rightarrow\infty,
	\end{equation}
	for fixed integers $a_r$, we find
	\begin{equation}
		\omega^2_{NA}
		\longrightarrow
		\omega_A^2,
		\qquad N\rightarrow\infty.
	\end{equation}
This establishes the pointwise convergence of the regularized frequencies to the continuum frequencies for each fixed Fourier mode. Consequently, for fixed $A$ and fixed occupation numbers, the corresponding modal energy contributions also converge. However, this mode-by-mode convergence does not by itself imply the convergence of the full ordered spectra. In particular, different Fourier modes may give rise to degenerate eigenvalues, and these degeneracies may split or recombine as $N$ varies. Therefore, the pointwise convergence of the frequencies does not provide direct control over the ordering and multiplicity of the eigenvalues. To address this issue, we now present a separate spectral argument based on the min-max theorem.
	
Let the set $\Lambda$ be
	\begin{eqnarray}
		\Lambda
		:=
		\{
		(p_1,p_2),\ 
		p_i=0,1,\dots,N-1;\
		i=1,2,\
		\mbox{excluding}\ (0,0):
		p_1^2+p_2^2<p^2
		\},
	\end{eqnarray}
	where $p$ is a fixed real number that may be arbitrarily large.
	
	We define
	\begin{eqnarray}
		H_N
		:=
		H^{scl}_{reg}\Bigg|_{A=(a_1,a_2)\in\Lambda},
		\qquad
		H_{\infty}
		:=
		H^{scl}\Bigg|_{A\in\Lambda}.
	\end{eqnarray}
	We also introduce
	\begin{eqnarray}
		\hat{H}_{\infty}
		=
		H^{scl}\Bigg|_{A\in\Lambda}
		+
		\sum_{A\in\Lambda}
		\omega_A^2 X^{mA}X_{mA},
	\end{eqnarray}
	as an auxiliary operator. It satisfies
	\begin{equation}
		\hat{H}_{\infty}>H_{\infty}.
	\end{equation}
	The common domain of $H_N$, $H_{\infty}$, and $\hat{H}_{\infty}$ is defined as
	\begin{eqnarray}
		D
		=
		\left\{
		\Psi\in L^2:
		\Psi\in H^1
		\ \mbox{with}\
		\left\langle
		\Psi,
		\left(\sum_{A\in\Lambda}
		\omega_A^2 X^{mA}X_{mA}\right)\Psi\
		\right\rangle
		<\infty
		\right\}.
	\end{eqnarray}
	The operators $H_N$, $H_{\infty}$, and $\hat{H}_{\infty}$ are self-adjoint, positive and have compact resolvent. Their eigenvalues $E_{Nk}$, $E_{\infty k}$, and $\hat{E}_{\infty k}$ are ordered as follows:
	\begin{eqnarray}
		&& E_{N1} \leq E_{N2} \leq \dots \leq E_{Nk} \leq \dots, \\
		&& E_{\infty 1} \leq E_{\infty 2} \leq \dots \leq E_{\infty k} \leq \dots, \\
		&& \hat{E}_{\infty 1} \leq \hat{E}_{\infty 2} \leq \dots
		\leq \hat{E}_{\infty k} \leq \dots,
	\end{eqnarray}
	counting multiplicities. Moreover, since $\hat{H}_{\infty}>H_{\infty}$,
	\begin{eqnarray}
		\hat{E}_{\infty k}>E_{\infty k}.
	\end{eqnarray}
\begin{lemma}
	Let $E$ be an arbitrarily large but fixed energy bound. Then
	\begin{eqnarray}
		E_{Nk}\longrightarrow E_{\infty k},
	\end{eqnarray}
	as $N\to\infty$, with eigenvalues counted according to multiplicity,
	for every $k\leq k_{\max}$, where $\hat{E}_{\infty k_{\max}}$
	is the largest eigenvalue of $\hat{H}$ strictly below $E$.
	
	The convergence is uniform over this range of $k$.
\end{lemma}
	
	\begin{proof}
		Using
		\begin{eqnarray}
			\Bigg|
			N^2\sin^2\!\left(\frac{\pi a_i}{N}\right)
			-
			(\pi a_i)^2
			\Bigg|
			\leq
			O\left(\frac{p^4}{N^2}\right),
		\end{eqnarray}
		it follows that, 
		
		\begin{align}
	 \langle\Psi, H_\infty-H_N \Psi\rangle =\langle\Psi, \sum_{A} \left(\frac{\omega_{A}-\omega_{NA}}{\omega_A}\right)\left((\omega_{A}+\omega_{NA})X^{\hat{m}A}X_{\hat{m}A}+i\omega_{A}(\Theta^T_{1A}\Theta^A_1-\Theta^T_{2A}\Theta^A_2)\right)\Psi\rangle,
		\end{align}
		in the large $N$ limit 
	\begin{align}
		\langle\Psi, H_\infty-H_N \Psi\rangle = \epsilon_N \langle\Psi, \hat{V}_{\infty} \Psi\rangle,
	\end{align}
	where $\hat{V}_{\infty}$ is the potential of $\hat{H}_{\infty}$. Since the kinematic and topological terms are positive we have
		\begin{eqnarray}
			\vert\langle\Psi,(H_\infty-H_N)\Psi\rangle\vert
			\leq
			\epsilon_N
			\langle\Psi,\hat{H}_{\infty}\Psi\rangle,
			\qquad
			\epsilon_N
			=
			O\left(\frac{p^4}{N^2}\right),
		\end{eqnarray}
		for sufficiently large $N$ and $\Psi\in D$. \newline
		
		For any finite dimensional space $V$
		\begin{eqnarray}
			\langle\Psi,\hat{H}_\infty\Psi\rangle\le \hat{E}_{\infty m}
		\end{eqnarray}
		where $\Psi_m$ is the eigenfunction of $\hat{H}_{\infty}$, with maximun eigenvalue $\hat{E}_{\infty m}$ in $V$. Notice that any finite dimensional space $V$ can be generates from a basis of eigenfunctions of $\hat{H}_\infty$.
		Hence, under the assumption of the Lemma 
		\begin{eqnarray}
			\vert\langle\Psi,(H_N-H_\infty)\Psi\rangle\vert
			\leq
			\epsilon_N E,
		\end{eqnarray}
		then
		\begin{eqnarray}
			\langle\Psi,H_\infty\Psi\rangle
			-
			\epsilon_N E\leq  \langle\Psi,H_N\Psi\rangle\leq \langle\Psi, H_\infty\Psi\rangle+\epsilon_N E.
		\end{eqnarray}
		The min-max theorem then implies
		\begin{eqnarray}
			\vert E_N-E_\infty\vert
			\leq
			\epsilon_N E,
		\end{eqnarray}
		for large enough $N$. Consequently,
		\begin{eqnarray}
			E_{Nk}\longrightarrow  E_{\infty k} 
		\end{eqnarray}
		uniformly, since $\epsilon_N$ does not depend on $k$, for all $k$. Such that,
		\begin{eqnarray}
			E_{\infty k}\leq E
		\end{eqnarray}
		since $E_{\infty k}\leq \hat{E}_{\infty k}$.
	\end{proof}
	
	\begin{remark}
		The introduction of $\Lambda$ in addition to $E$ is necessary since a bound on the frequencies does not imply a bound on the values of $a_1$ and $a_2$. In fact,
		\begin{eqnarray}
			\sin\left(\frac{\pi a}{N}\right)<\frac{k}{N}
		\end{eqnarray}
		admits solutions of the form $a=N-m$, with $m<k$. Hence, there are two branches involved in the semiclassical analysis. The additional condition $|a_i|<p$, $i=1,2$, eliminates the branch $N-m$, with $m<k$. This is a property of the spectrum of the semiclassical model $H_{reg}^{scl}$, which is invariant under $a_i\rightarrow N-a_i$. This symmetry is not expected when considering the exact $H^{scl}$.
	\end{remark}
	
	The lemma provides an additional spectral result that does not follow directly from the preceding mode-by-mode convergence. By applying the min-max theorem, it controls the ordered eigenvalues and their multiplicities uniformly below a fixed energy bound, thereby taking into account possible degeneracies. Therefore, within the semiclassical approximation considered here, the large-$N$ limit of the regularized Hamiltonian reproduces the continuum spectrum.
	
	\section{Conclusions}
	
	The large-$N$ limit of the matrix-regularized supermembrane remains one of the central open problems in the formulation of the theory. This issue is directly related to the question of whether the finite-dimensional matrix model provides a well-defined approximation to the continuum supermembrane theory. In this work we have addressed this problem at the semiclassical level for the supermembrane with central charge.
	
	We have established the large-$N$ limit of the semiclassical regularized supermembrane and demonstrated its convergence to the corresponding unregularized semiclassical description. Both Hamiltonians reduce to supersymmetric harmonic oscillators. The convergence of the regularized frequencies to their continuum counterparts implies pointwise convergence of individual eigenvalues. In addition, we prove a lemma based on the min--max principle that establishes convergence of the full spectra, with eigenvalues counted according to multiplicity.
	
	This result extends the pointwise convergence previously established by the authors in the bosonic case \cite{mpgm17} to the full supersymmetric spectrum, connecting the regularized M2-brane model in the large-$N$ limit to its continuum semiclassical description. The supersymmetric extension is also relevant because the fermionic contribution cancels the bosonic zero-point energy mode by mode. Consequently, the divergent constant arising from the infinite sum over oscillator modes in the purely bosonic description is absent from the supersymmetric semiclassical Hamiltonian.
		
	Furthermore, we have introduced the treatment of the global APD constraints at the regularized level. These constraints carry non-trivial topological information associated with the compact sector of the theory. To the best of our knowledge, in previous treatments of the matrix regularization this global contribution had not been explicitly incorporated, leading to an incomplete description of the compact momenta components. In the present work we have shown that the global APD constraint determines the zero modes of the momenta conjugate to the compact directions in terms of the regularized APD degrees of freedom. Hence, these zero modes are not free parameters, in contrast with the zero modes associated with the non-compact directions.
	
	The results obtained here provide an explicit test of the correspondence between the regularized and the continuum descriptions in a sector where the spectrum can be computed. Nevertheless, extending this correspondence beyond the semiclassical approximation remains a highly non-trivial problem. A natural next step is therefore to analyze whether the large-$N$ convergence persists for the full interacting Hamiltonian, where the interaction terms and the structure of the constraints must be treated without relying on the quadratic approximation.




\medskip
\textbf{Acknowledgements} \par 

MPGM has been  partially supported by the PID2024-155685NB-C21 MCI Spanish Grants and by the University of La Rioja project REGI2025/41.  P. L. is funded by ANID/Fondecyt Postdoctorado 2025/3250504. AR and  want to thank to protect MATH-AMSUD 240048 and the Scientific Research Computing Institute of the University of La Rioja (SCRIUR), Spain.

\medskip 
\textbf{Conflic of Interest}
The authors declare no conflicts of interest.

\medskip

%
\bibliographystyle{MSP}

\bibliography{example}


\end{document}